\documentclass[letters,fleqn,usenatbib]{mnras}

\usepackage{mathptmx}
\usepackage{pdflscape}

\usepackage[T1]{fontenc}
\usepackage{ae,aecompl}

\usepackage{lipsum}

\usepackage{graphicx}	
\usepackage{amsmath}	
\usepackage{amssymb}	
\usepackage{multirow} 
\usepackage{subfigure}
\usepackage{threeparttable}
\usepackage{color}
\usepackage{colortbl}

\usepackage[cmyk]{xcolor}
\definecolor{Gray}{gray}{0.9}

\newcommand{\citefoot}[2]{\citeauthor{#1}~(\citeyear[#2]{#1})}

\title[CO $v=1, J=3-2$ detected towards five AGB stars]{ALMA observations of the vibrationally-excited rotational CO transition $v=1, J=3-2$ towards five AGB stars}

   \author[T. Khouri et al.]{T. Khouri,
           $^{1}$\thanks{\it Email: theokhouri@gmail.com}
           W. H. T. Vlemmings,$^{1}$
           S. Ramstedt,$^{2}$
           R. Lombaert,$^{1}$
           \newauthor
           M. Maercker$^{1}$
           and E. De Beck$^{1}$
\\
$^1$Department of Earth and Space Sciences, Chalmers University of Technology, Onsala Space Observatory, 439 92 Onsala, Sweden \\ 
$^2$Department of Physics and Astronomy, Uppsala University, Box 516, 75120 Uppsala, Sweden 
}

\date{Accepted 2016 August 9. Received 2016 August 8; in original form 2016 July 14}

\pubyear{2016}

\begin{document}
\label{firstpage}
\pagerange{\pageref{firstpage}--\pageref{lastpage}}
\maketitle

\begin{abstract}
We report the serendipitous detection with ALMA of the vibrationally-excited pure-rotational CO transition $v=1, J=3-2$ towards five asymptotic
giant branch (AGB) stars, $o$~Cet, R~Aqr, R~Scl, W~Aql, and $\pi^1$~Gru. The observed lines are formed in the poorly-understood region
located between the stellar surface and the region where the wind starts, the so-called warm molecular layer. We successfully reproduce the observed lines
profiles using a simple model. We constrain the extents, densities, and kinematics of
the region where the lines are produced. R~Aqr and R~Scl show inverse P-Cygni line profiles which indicate infall of material onto the stars.
The line profiles of $o$~Cet and R~Scl show variability.
The serendipitous detection towards these five sources shows that vibrationally-excited rotational lines
can be observed towards a large number of nearby AGB stars using ALMA. This opens a new possibility for the study of the innermost regions of
AGB circumstellar envelopes.
\end{abstract}

\begin{keywords}
stars: AGB and post-AGB -- (stars:) circumstellar matter
\end{keywords}



\section{Introduction}

During the asymptotic giant branch (AGB), low- and intermediate-mass stars present strong winds thought to be
driven by radiation pressure on dust grains. Dust formation is made possible
by the action of stellar pulsations and convection, which levitate gas to high distances from the star where temperatures are low enough.
This process creates a region with enhanced gas densities, referred to as warm molecular layer \citep[WML,][]{Tsuji1997}.
Although models for the dynamical extended atmospheres of AGB stars
have advanced significantly in recent years \citep[see, e.g.,][and references therein]{Bladh2015,Hoefner2016}, {our understanding of this complex region is still incomplete,
and calculating the mass-loss rate, $\dot{M}$, of AGB stars from first principles is not possible presently.}
In this context, empirically characterising the properties of the gas from which dust forms is paramount. This region is usually
studied through spectroscopic observations of
molecular bands \citep[see, e.g.,][]{Tsuji1997,Mennesson2002,Perrin2004,Ohnaka2012} and of resolved molecular
lines \citep[see, e.g.,][]{Hinkle1982, Lebzelter2005, Nowotny2010} in the near-infrared. By using different molecules and transitions, authors are able to constrain the kinematics and the column densities
of gas in the WML, even as a function of the distance from the star.

We report the serendipitous detection of the CO~${v=1, J=3-2}$ transition using ALMA towards all five AGB stars for which data was available to us: R~Aqr, R~Scl, $o$~Cet, W~Aql, and $\pi^1$~Gru. 
This transition has an excitation energy of $3117~{\rm K}=2686~{\rm eV}=2167~{\rm cm}^{-1}$ and
probes gas in the WML and in the dust formation region.
Since CO is a well-understood molecule, the optical depth in this line is not high,
and this transition does not show maser action, these observations are ideal to study the conditions of the gas in the WML.
The observed lines offer snapshots of the gas kinematics in the WMLs of the five stars.
To our knowledge, the only previous detection of a vibrationally-excited pure-rotational CO transition from an AGB star
is towards the C-rich star IRC~+10216 with the Submillimeter Array \citep{Patel2009}.

\section{Observations}

The observations of $o$~Cet, W~Aql and $\pi^1$~Gru were acquired in mosaic mode and those of R~Aqr and R~Scl were obtained in a single pointing.
The observations of $o$~Cet were presented by \cite{Ramstedt2014a}.
The CO~${v=1, J=3-2}$ transition at 342.647~GHz \citep{Gendriesch2009} was detected in the observed spectral windows 0.937~GHZ wide and centred at 342.883~GHz for R~Scl
and 1.875~GHz wide and centred at 343.30~GHz for the other four sources. CO has no other vibrationally-excited transitions in the observed spectral range.
Table~\ref{tab:obs} summarises the observations. 
The emission region is spatially unresolved for
the five sources, consistent with the expected line-formation
region with $r < 10~R_\star$ (see Section~\ref{sec:disc}).
{The CO~${v=1, J=3-2}$ transition show clear inverse P-Cygni profiles towards
R~Aqr and R~Scl}, which suggests material falling back to the stars.

$o$~Cet, W~Aql, and $\pi^{1}$~Gru were observed in two epochs.
The signal-to-noise ratio (S/N) in the observations of $o$~Cet is high enough that variability can be seen.
For W~Aql, the lines observed 3.5 weeks apart do not differ significantly. The line is not detected towards $\pi^{1}$~Gru
in the individual epochs because of the low S/N. Therefore, we have averaged the two epochs of observation of W~Aql and
of $\pi^{1}$~Gru, but we analyse those of $o$~Cet individually. R~Scl was observed three times, twice at maximum-light phase and once at minimum-light phase.
The two line profiles seen during maximum-light phase are consistent within the uncertainties and were averaged to increase the S/N, while that
obtained at minimum-light phase shows a significantly different line profile and was analysed separately.

\section{Modelling}
\label{sec:models}

For simplicity, we divide our model circumstellar envelopes in two regions: the WML and the outflow.
We define the interface between these two region as the radius
at which the acceleration of the wind starts, $R_\circ$.
Densities of $\gtrsim 10^{10}$~cm$^{-3}$  are required for the rotational levels in the first vibrational state to be populated through collisions \citep{Woitke1999}.
In the WML, the densities are expected to be of this order or higher but when the outflow starts the densities quickly drop well below this value for the five sources.
A possible exception is the inner region of the outflow of R~Aqr, which has a high, but rather uncertain, $\dot{M}$ (see below).

\vspace{0.2cm}
{\bf \noindent Emission from the outflow:}
{The low gas densities in the outflow cause the level populations of the $v=1$ level to be mainly excited by the radiation field at 4.6\,$\mu$m.
To estimate the strength of the emission expected from $r > R_\circ$,
we used GASTRoNOoM \citep{Decin2006} to calculate non-LTE spherically-symmetric radiative transfer}
models for the outflows of the five sources. We assume that the number density at a given radius $r$ is given
by ${n(r) = \dot{M}/ (4\pi r^2 \upsilon(r))}$, where ${\upsilon(r) = \upsilon_\circ +
(\upsilon_\infty - \upsilon_\circ) (1-R_\circ/r)^{1.0}}$ is the expansion velocity at a given radius, with $\upsilon_\infty$ being
the terminal expansion velocity of the outflow.
We adopt an initial expansion velocity, $\upsilon(R_\circ) = \upsilon_\circ$, of 2~km/s. This is a conservative lower limit on the sound speed at
$R_\circ$ and, hence, gives an upper limit on the gas densities at the start of the outflow.
For $\dot{M}$ and $\upsilon_\infty$ we use values available from literature (see Table~\ref{tab:obs}).
We use a stellar temperature of 3000~K.

We find that the predicted line strengths of transition ${v=1, J=3-2}$ of CO for the outflows of $o$~Cet, W~Aql, and $\pi^1$~Gru are
weaker than the observed emission by factors of roughly 50, 20, and 10, respectively. Hence, the WML dominates emission in this line for these sources.
For R~Aqr and R~Scl, we find line strengths that are comparable to those observed for these stars. However, the mass-loss rates 
of these two sources are rather uncertain. For R~Aqr, deriving $\dot{M}$ is difficult because the system is a very close binary.
\cite{Bujarrabal2010} obtained the value we use by assuming that the emission region of CO transition $v=0, J=2-1$
is 0''.09 in radius, based on the distance between R~Aqr and its companion. However, our observations of CO $v=0, J=3-2$ show an asymmetric emission region which is up to about ten times
larger in volume than the one considered by \citeauthor{Bujarrabal2010} (Ramstedt et al., {\em in prep.}). Therefore, the mass-loss rate derived by them is probably over-estimated.
For R~Scl, $\dot{M}$ has been steadily decreasing since the onset of a thermal pulse $\sim 2000$~yr ago \citep{Maercker2016}. This makes the derivation
of the present-day value difficult.

Given the small contribution of emission from the outflowing gas to the observed line profile for $o$~Cet, W~Aql, and $\pi^1$~Gru and the large uncertainty
on the $\dot{M}$ of R~Aqr and R~Scl, we have
not considered emission from the outflow when fitting the observations.
We discuss the possible effects of this approximation in Section~\ref{sec:fits}.

\vspace{0.2cm}
{\bf \noindent The warm molecular layer:}
To fit the observed lines, we calculated radiative-transfer models consisting of a spherical-symmetric
WML that extends radially from the stellar surface, $r=R_\star$, to the edge of the outflow, $r=R_\circ$.
We opt for a simple model in which the density, $n_{\rm WML}$, and excitation temperature of CO, $T^{\rm Vib}_{\rm WML}$,
are constant throughout the WML.
This is motivated by our lack of understanding of the density distribution and variability of this complex region
and by the fact that such simple models produce good fits to the observations.
We calculated models with $T^{\rm Vib}_{\rm WML}$ equal to 1000~K, 1500~K, and 2000~K.
The velocity field of the gas in the WML is described by a velocity dispersion,
$\upsilon^{\rm Disp}_{\rm WML}$, and a global radial velocity, $\upsilon_{\rm WML}$.
$R_\circ$, $n_{\rm WML}$, $\upsilon^{\rm Disp}_{\rm WML}$, and $\upsilon_{\rm WML}$
were varied to fit the observations.

We assume the stellar temperatures to be 3000~K and we use the stellar continuum flux density, $S_\star$, measured by ALMA to constrain the
stellar radius through
\begin{equation}
R_\star = \sqrt{\frac{S_\star(342.65~{\rm GHz}) \, d_\star^2}{\pi \, B(342.65~{\rm GHz}, 3000~K)}} ,
\end{equation}
where $B\left (342.65~{\rm GHz}, 3000~K\right )$ is the Planck function for a 3000~K black-body evaluated at 342.65~GHz and $d_\star$ is the distance to the star.
The model line profiles were constructed by calculating the emerging flux density
\begin{equation}
S\left(\nu\right) = \frac{2\pi}{d^2}\int^{R_\circ}_0 I_{p}\left(\nu\right) p dp
\end{equation}
at a given frequency $\nu$. The intensity at $\nu$ was calculated for a number of lines of sight that intersect the WML, with impact parameter $p$ ranging
from zero to $R_\circ$.
using the equation
\begin{equation}
I_p(\nu) = I_{p}^{\circ}(\nu) e^{-\tau_\nu} + B(\nu,T_{\rm WML}^{\rm Vib})[1-e^{-\tau_\nu}].
\end{equation}
The initial intensity, $I_{p}^{\circ}(\nu)$, was set to zero for $p>R_\star$ and to $B(342.65~{\rm GHz}, 3000~K)$ for $p\leq R_\star$.
We have not considered the presence of a companion star
in our calculations. This approach is justified because our calculations show that
the $v=1, J=3-2$ CO line is formed very close to the star ($r \lesssim 3~R_\star$) and maser emission lines formed
at similar distances observed towards close binaries AGB stars, such as R~Aqr, do not show distinct features from those observed towards
single AGB stars \citep[see, e.g.,][]{Kamohara2010}. 
{We note, however, that the five AGB stars discussed here are peculiar in some way because of interactions with their companions.
Future studies are needed to show whether the line profiles they present differ from those seen towards single AGB stars.}

\section{Discussion}
\label{sec:disc}

\begin{figure}
   \centering
   \includegraphics[width= 7.3cm]{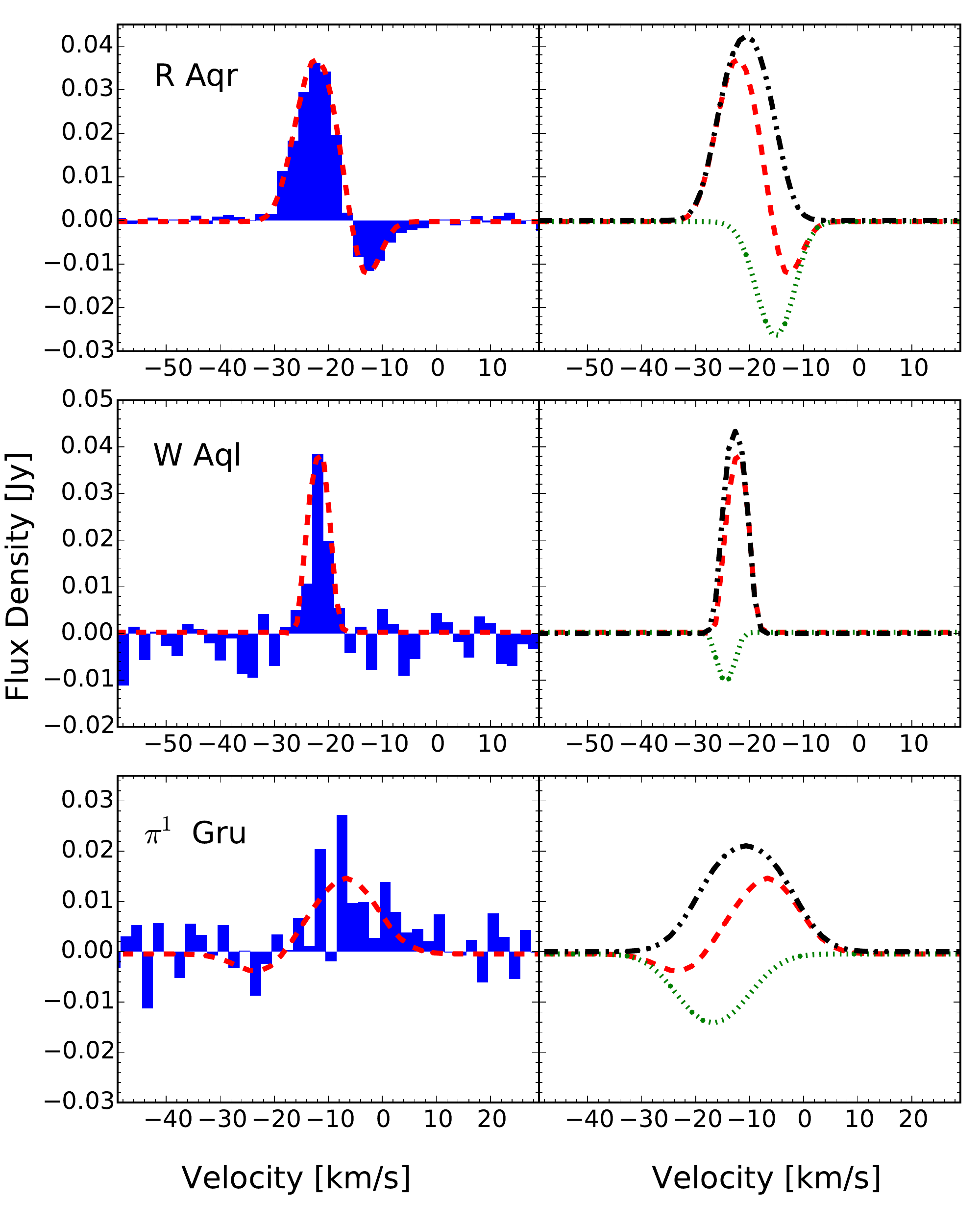}
      \caption{{\em Left panels}: best fit models with $T^{\rm Vib}_{\rm WML} = 1500$~K (dashed red line) compared to the observed line profiles of R~Aqr, W~Aql, $\pi^1$~Gru (blue histogram).
      {\em Right panels}: best fit models (dashed red line, same as in left panels)
      compared to the contribution from lines of sight that intersect the star 
      (dotted green line) and
      that do not intersect the star (dotted-dashed black line).}
         \label{fig:modelsa}
   \end{figure}

\begin{figure}
   \centering
   \includegraphics[width= 7.3cm]{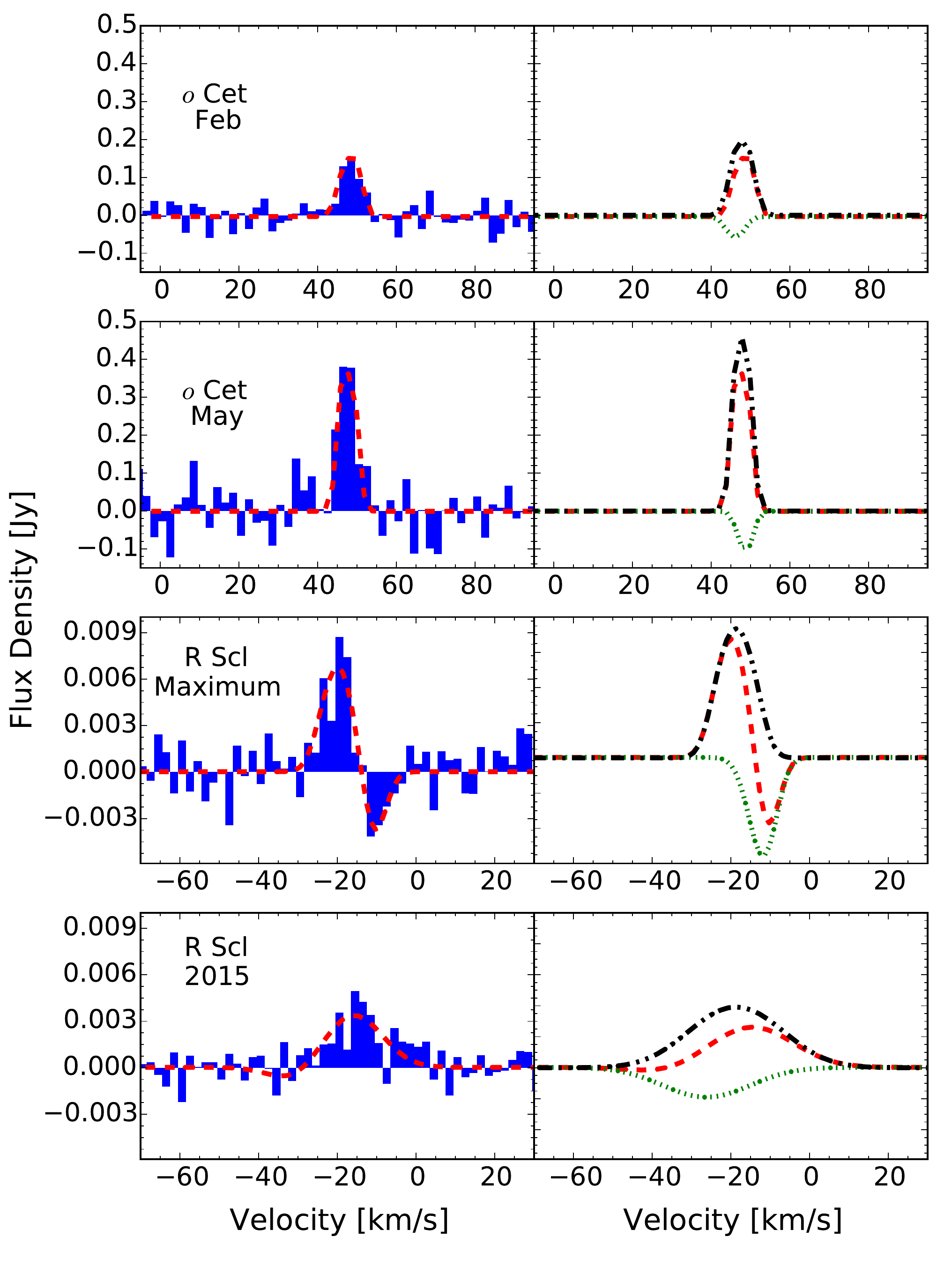}
      \caption{{\em Left panels}: best fit models with $T^{\rm Vib}_{\rm WML} = 1500$~K compared to the observed line profiles of $o$~Cet and R~Scl in two epochs for each source.
      {\em Right panels}: best fit models (dashed red line, same as shown in the left panels)
      compared to the contribution from lines of sight with $p \leq R_\star$ (dotted green line) and
      $p>R_\star$ (dotted-dashed black line).}
         \label{fig:modelsb}
   \end{figure}

{The values of the best-fit parameters are given in Table~\ref{tab:models} and the fits to the data are shown in Figs.~\ref{fig:modelsa} and \ref{fig:modelsb}.}
For the inverse P-Cygni line profiles, we obtained the $1\sigma$ uncertainties from mapping the $\chi^{2}$ around the best model given in Table~\ref{tab:models}.
For $o$~Cet and W~Aql, the errors were derived from the uncertainty on the line fluxes.
All the uncertainties concern only the fitting procedure, and the systematic errors from our model assumptions can be
larger than those.

{In our models for R~Aqr, R~Scl, and $\pi^1$~Gru, the maximum optical depths for the different impact parameters and velocities are $\lesssim 1$.
Thanks to this, the masses we derive for the WMLs of these stars do not dependent strongly on the assumed morphology and density profile. For
$o$~Cet and W~Aql, the maximum optical depths reach $\sim 3.5$ for the models given in Table~\ref{tab:models}
(which have the maximum possible optical depths and gas masses, see below).
The extent of the WML derived from our models are comparable to those derived for 14 AGB stars from measurements of angular sizes
in the $K'$ and $L'$ bands \citep{Mennesson2002}. 
Our results are also in good agreement with those of \cite{Perrin2004} who derive radii of about 2.2\,$R_\star$ and temperatures ranging from 1500-2100~K
for the WMLs of a sample of five Mira stars by using CO and H$_2$O lines from near-infrared spectra.}

\subsection{Model fits}
\label{sec:fits}

{\bf $o$~Cet and W~Aql:}
We constrained $\upsilon_{\rm WML}$ by requiring that the line position matched the observations given the $\upsilon_{LSR}$ (see Table~\ref{tab:models}). This leads to
low values of $\upsilon_{\rm WML}$ for the two sources. Since the $\upsilon_{LSR}$ is usually not known with better precision than $\sim 1\,$km\,s$^{-1}$,
the values of $\upsilon_{\rm WML}$ we derive are also uncertain by a similar amount. $\upsilon^{\rm Disp}_{\rm WML}$ was then determined by fitting the width of the observed
line. The other three free parameters, $n_{\rm WML}$, $R_\circ$, and $T^{\rm Vib}_{\rm WML}$, show degeneracy for these two stars.
Nonetheless, there is a limiting value of $R_\circ$ below which a good fit to the observed line cannot be found for a given value of $T^{\rm Vib}_{\rm WML}$.
This is because the line would become too optically thick before the observed flux is reached. 
Hence in Table~\ref{tab:models}, we give the minimum value of $R_\circ$ and the maximum value of $n_{\rm WML}$
for which a good fit can be obtained for each given $T^{\rm Vib}_{\rm WML}$.
If the other parameters are kept fixed, equivalent fits can be obtained by increasing $R_\circ$
with respect to the values given in Table~\ref{tab:models} and decreasing
$n_{\rm WML}$ accordingly. This causes the optical depth of the line
to decrease and, hence, such models require lower gas masses in the WML.
Since collisional
excitation of the $v=1, J=3$ level of CO is no longer efficient for $n \sim 10^9$\,cm$^{-3}$, the size of the WML of these two stars is probably $\lesssim 10$\,AU. From our models, we can then set
upper limits on the fraction of the mass of the WML lost per pulsation cycle ($\frac{1}{M_{\rm WML}}\frac{dM_{\rm WML}}{dP}$) to be $\lesssim 1.0\%$ for $o$~Cet and
$\lesssim 7.5\%$ for W~Aql (see Table \ref{tab:models}).

{\bf \noindent R~Aqr and R~Scl:} Our models require the gas
in the WMLs to be falling back to the stars, as expected from the line profiles.
The parameters we derive for the WML of these two stars are better constrained than those for $o$~Cet and W~Aql.
This is because the amount of absorption observed becomes smaller when we increase either $T^{\rm Vib}_{\rm WML}$ or $R_\circ$,
while keeping the other parameters fixed. Therefore, there is no degeneracy between $R_\circ$ and $n_{\rm WML}$ and
only one combination of these two parameters fits the observed line for a
given value of $T^{\rm Vib}_{\rm WML}$. 
We note that we have not included emission from the outflow in our modelling although this might be important for these two sources. Our models show
that emission from the outflow would be centred on the $\upsilon_{\rm LSR}$ and that a small amount of red-shifted absorption could be produced. Taking this
emission into account would require
the value of $R_\circ$ to decrease and the value of $n_{\rm WML}$ to increase for a given value of $T^{\rm Vib}_{\rm WML}$.
As an example, the model for R~Aqr with $\dot{M}_{\rm gas} = 9\times10^{-6}$\,M$_\odot$\,yr$^{-1}$ and $T^{\rm Vib}_{\rm WML}=1500$~K would
require $R_\circ$ to decrease from 1.45~$R_\star$ to 1.35~$R_\star$ and $n_{\rm WML}$ to increase from
$1.1\times10^{11}$~cm$^{-3}$ to $1.3 \times 10^{11}$~cm$^{-3}$.

{\bf \noindent $\pi^1$~Gru: } our models are only able to reproduce the observed red-shift with respect to the $\upsilon_{\rm LSR}$ if absorption is present in the blue-shifted part of the line.
The S/N is not sufficiently high for such absorption to be seen, however.
Alternative explanations to the asymmetric line profile are possible if asymmetries
in the envelope and/or velocity field are invoked. We also note that since we have averaged two datasets taken approximately 1.5 years ($\sim 3.25$ pulsation periods) apart,
the line profile we model might not be representative of either epoch.

\subsection{Variability}
\label{sec:variability}

{\bf \noindent $o$~Cet:} the two observations of $o$~Cet are separated by 5.5 weeks and
the peak flux density increases from 0.15~Jy close to minimum-light phase to 0.40~Jy
close to maximum-light phase.
This variation can be reproduced by an increase in $R_\circ$, $n_{\rm WML}$, or $T^{\rm Vib}_{\rm WML}$.
Changes in $R_\circ$ larger than 0.2~AU between the two epochs are unlikely because these would imply
average expansion velocities ${\geq10}$~km\,s$^{-1}$, which are not consistent with the observations.
Regarding $n_{\rm WML}$, an increase
by a factor $> 5$ between the two epochs would be required to explain the increase in line strength.
This seems unlikely given the large extent of the WML.
An increase of $T^{\rm Vib}_{\rm WML}$ of $\sim 1000$~K would cause the
line strength to increase by the required amount without requiring a significant change of $R_\circ$ or $n_{\rm WML}$.
In this case, $T^{\rm Vib}_{\rm WML}$ would be higher close to maximum light phase.
Hence, an increase
in $T^{\rm Vib}_{\rm WML}$ is the most probable cause to the observed line strength increase.

{\bf \noindent R~Scl:}  the observed line shapes are not
strongly constrained, specially in the observation from July 2015.
Therefore, we cannot draw strong conclusions on the variability of the WML of R~Scl.
Nonetheless, our models indicate that from maximum- to minimum-light phase the $\upsilon_{\rm WML}$ changes from negative to positive
and $\upsilon^{\rm Disp}_{\rm WML}$ increases.

\subsection{Comparison between the five sources}
\label{sec:comp}

Our results suggest that at most about 10\% of the mass in the WML is lost in each pulsation cycle. This would imply that most of the mass
lifted by pulsations falls back to the stars, or remains in the WMLs for at least about 10 pulsation cycles.
Interestingly, for $o$~Cet only $\lesssim 1\%$ of the mass in the WML seems to be lost per pulsation cycle, a lower fraction than for the other stars.
This difference might not be real if the $\dot{M}_{\rm gas}$ values we use are off by about a factor of ten.

The two inverse P-Cygni profiles we detect are clearly seen, while the (tentative) regular P-Cygni profiles are weak and show no
absorption above the noise level. This could be caused by the different S/N between the observations but we note that the emission component of the
inverse P-Cygni profile observed towards R~Scl is detected with a S/N comparable to the emission component seen towards R~Scl
in 2015 and $\pi^1$~Gru. The reason the regular P-Cygni profiles appear to be weaker is because their emission components are broader than those of the
inverse P-Cygni profiles. {Both the higher value of the FWHM and the lack of absorption could be explained by a larger value of the
$\upsilon^{\rm Disp}_{\rm WML}$ for the regular P-Cygni profiles than for the inverse P-Cygni profiles.}

{Finally, our model fits suggest that the sources which show emission peaking at $\upsilon_{\rm LSR}$, , W~Aql and $o$~Cet, have
a relatively more extended WML than those which do not, R~Aqr, R~Scl and $\pi^1$~Gru.} This is because
smaller values of $R_\circ$ increase the strength of the absorption relatively to the emission.
Hence, these difference in the derived values of $R_\circ$ seem to reflect intrinsic differences between the circumstellar environment of these stars at the times of the observations.

The high (serendipitous) detection rate we report (5 out of 5 sources) shows that the sensitivity of ALMA allows to detect
emission in pure-rotational lines in the first vibrational excited state for AGB stars within at least $\sim 400$\,pcs.
{Our models show that $v=1$rotational transitions should be stronger the higher the rotational quantum number, but transition
$v=1, J=3-2$ is probably the easiest to detect from the ground nonetheless.}
Observations of these transitions set important constraints to the properties of AGB WMLs.
This opens an exciting new possibility for the study of the innermost regions of AGB circumstellar envelopes.

\section*{Acknowledgements}
\addcontentsline{toc}{section}{Acknowledgements}
This paper makes use of ALMA data from projects Nos. ADS/ JAO.ALMA\#2012.1.00524.S and ADS/ JAO.ALMA\#2012.1.00097.S.
ALMA is a partnership of ESO (representing its member states), the NSF (USA) and NINS (Japan), together with the NRC (Canada),
 NSC and ASIAA (Taiwan) and KASI (Republic of Korea), in cooperation with the Republic of Chile. The Joint ALMA Observatory is operated by ESO, AUI/NRAO and NAOJ.
This work is supported by VR. W.V. acknowledges support from ERC consolidator grant 614264.
M.M. has received funding from the People Programme (Marie Curie Actions) of the EU's FP7 (FP7/2007-2013)
under REA grant agreement No. 623898.11.

\begin{landscape}
\begin{table}
\begin{center}
\small
\caption[]{The distances ($d$), periods ($P$), maximum expansion velocities ($\upsilon_\infty$),
gas mass-loss rates ($\dot{M}_{\rm gas}$), and dates and {pulsation phases of the observations
($\phi$, obtained from data from the American association of variable stars observers, http://www.aavso.org)} are shown for all sources.
The continuum fluxes ($S_\star$), line peak fluxes ($S_{\rm peak}$), integrated line intensities ($\int S d\upsilon$), full-width at half maximum of the lines (FWHM),
root-mean squared noise of the observations ($rms$),
and smaller dimension of the beam size all obtained from the ALMA observations are also given.
For the inverse P-Cygni profiles, we give $S_{\rm peak}$ and $\int S d\upsilon$ of the absorption component as negative values.
The distances were retrieved from \citefoot{vanLeeuwen2007}{$o$~Cet}, \citefoot{Whitelock2008}{R~Scl and W~Aql}, \citefoot{Min2014}{R~Aqr}, and \citefoot{Mayer2014}{$\pi^1$~Gru}.
The $\dot{M}_{\rm gas}$ are those given by \citefoot{DeBeck2010}{$o$~Cet and $\pi^1$~Gru}, \citefoot{Maercker2016}{R~Scl}, \citefoot{Danilovich2014}{W~Aql}, and \citefoot{Bujarrabal2010}{R~Aqr}.
$P$ and $\upsilon_\infty$ were retrieved from \citefoot{Bujarrabal2010}{} for R~Aqr and from \citefoot{DeBeck2010}{} for the other sources.}
\begin{tabular}{ccccccccccccc}
\hline \noalign {\smallskip}
\label{tab:obs}
Source & $d$ & $P$ & $\upsilon_\infty$ & $\dot{M}$ & Date & $\phi$ &  $S_\star$ & $S_{\rm peak}$ & $\int S d\upsilon$ & FWHM & $rms$ & Beam \\
& [pcs] & [day] & [km s$^{-1}$] & [M$_\odot$\,yr$^{-1}$]  & [yy-mm-dd] &  & [mJy] & [mJy] & [Jy km/s] & [km/s] & [mJy] & [arcsec]\\
\hline \noalign {\smallskip}
$o$~Cet & 107 & 331 & 8.1 &  $2.5\times10^{-7}$ & 14-02-25 & 0.70 &160 & 160 & $460 \pm 120$ & 6.5 & 33 & 0.54 \\
$o$~Cet &  &  & & & 14-05-03 & 0.95 & 234 & 380 & $1200\pm250$ & 6.0 & 60 & 0.33 \\
\rowcolor{Gray}
R~Scl & 370 & 370 & 17.0 & $1 \times 10^{-6}$ & 2 epochs$^{\rm a}$ & 0.00/1.00 & \phantom{0}32 & 8/-4.6 & $30\pm5$/-$15\pm 5$  & $\sim 7 + 6.5$ & 1.6 & 0.15/0.70 \\
\rowcolor{Gray}
R~Scl &  &  & & & 15-07-24 & 1.55 & \phantom{0}23 & 5 & $29\pm5$ & $\sim$ 19 & 1.0 & 0.13 \\
W~Aql & 395 & 490 & 20.0 & $4\times 10^{-6}$ & 2 epochs$^{\rm a}$ & 0.40/0.45 & \phantom{0}26 & 40 & $96\pm15$ & 3.5 & 5.5 & 0.33/0.44 \\
\rowcolor{Gray}
R~Aqr & 218 & 387 & 20.0 & $9\times 10^{-6}$ & 15-05-21 & 0.05 & 110 & 36/-11.5 & $161\pm4$/-$55 \pm 4 $ & 8.5 / 6 & 1.2 & 0.34 \\
$\pi^1$~Gru & 163 & 195 & 30.0 & $8.5\times 10^{-7}$ & 2 epochs$^{\rm a}$ & & \phantom{0}66 & 27 & $115\pm25$ & $\sim 12$ & 5.5 & 0.47\\
\end{tabular}
\end{center}
\vspace{-0.3cm}
\begin{flushleft}
{a - the analysis was carried out on data obtained from the average of observations carried out on 2013-12-14
and 2014-12-25 for R~Scl, on 2014-03-20 and 2014-04-14 for W~Aql, and on 2013-10-08 and 2014-03-25 for $\pi^1$~Gru.}
\end{flushleft}
\end{table}

\begin{table}
\begin{center}
\footnotesize
\caption[]{Derived model parameters with associated uncertainty from the fitting procedure (see Section~\ref{sec:models}).
$\frac{1}{M_{\rm WML}}\frac{dM_{\rm WML}}{dP} $ is the fraction of the mass of the WML lost per pulsation cycle (see Table~\ref{tab:obs} for $\dot{M}$
and $P$).}
\begin{tabular}{c|cccccccccc}
\hline \noalign {\smallskip}
\label{tab:models}
Source & Date Obs. & $\upsilon_{\rm LSR}$ & $R_\star$ & $R_\circ$ & $n_\circ$ & $n_{\rm WML}$ & $\frac{1}{M_{\rm WML}}\frac{dM_{\rm WML}}{dP} $ & $T^{\rm Vib}_{\rm WML}$
& $\upsilon^{\rm Disp}_{\rm WML}$ & $\upsilon_{\rm WML}$ \\
 & [yy-mm-dd] & [km/s] & [AU] & [AU] & [cm$^{-3}$] & [cm$^{-3}$] & [\%] & [K] & [km/s] & [km/s] \\
\hline \noalign {\smallskip}
$o$~Cet  & 14-02-25 & +47.2 & 1.5 & $3.6\pm0.20$ & $6.5 \times 10^{8}$ & $7\pm2 \times 10^{10}$ & 0.3 & 1000 & $2.5\pm1.0$ & $2.0\pm1.0$ \\
 & &  &  & $3.0\pm 0.15$ & $9.3 \times 10^{8}$ & $9\pm 2 \times 10^{10}$ & 0.4 & 1500 & $2.5\pm1.0$ & $2.0\pm1.0$ \\
 & &  &  & $2.6\pm0.15$ & $1.2 \times 10^{9}$ & $1.3 \pm 0.3 \times 10^{11}$ & 0.5 & 2000 & $2.5\pm1.0$ & $2.0\pm1.0$ \\
\rowcolor{Gray}
$o$~Cet & 14-05-03 & +47.2 & 1.85 & $5.2\pm0.3$ & $3.1 \times 10^{8}$ & $4.5 \pm 1\times 10^{10}$ & 0.2 & 1000 & $2.0\pm1.0$ & $-1.0\pm1.0$ \\
\rowcolor{Gray}
 &  &  &  & $4.2\pm0.2$ & $4.8 \times 10^{8}$ & $7.0\pm 1\times 10^{10}$ & 0.2 & 1500 & $2.0\pm1.0$ & $-1.0\pm1.0$ \\
\rowcolor{Gray}
 &  &  &  &$3.6\pm0.2$ & $6.5 \times 10^{8}$ & $8.5 \pm 1 \times 10^{10}$ & 0.3 & 2000 & $2.0\pm1.0$ & $-1.0\pm1.0$ \\
R~Scl & 2 epochs$^{\rm a}$ & -19.0 & 2.35 & $4.0^{+0.7}_{-0.5}$ & $2.1 \times 10^{9}$ & { $1.5^{+2.0}_{-0.7} \times 10^{10}$} & 5.9 & 1000 & $3.5\pm1.5$ & { $-8.0\pm3.0$} \\ 
 &  & &  & $3.2^{+0.5}_{-0.3}$ & $3.6 \times 10^{9}$ & $5 \pm 4 \times 10^{10}$ & 4.5 & 1500 & $3.5\pm1.5$ & { $-8.0\pm3.0$} \\ 
 &  &  &  & $2.8\pm0.2$ & $4.2 \times 10^{9}$ & $1.0^{+1.5}_{-0.5} \times 10^{11}$ & 4.5 & 2000 & $3.5\pm1.5$ & { $-8.0\pm3.0$} \\ 
\rowcolor{Gray}
R~Scl & 15-07-24 & -19.0 & 2.0 & 3.6 & $2.6 \times 10^{9}$ & { $2.0 \times 10^{10}$} & 5.5 & 1000 & { 10.0} & { 10.0} \\  
\rowcolor{Gray}
 &  &  &  & { 2.8} & $4.3 \times 10^{9}$ & { $4.5 \times 10^{10}$} & 6.7 & 1500 & { 10.0} & { 10.0} \\ 
\rowcolor{Gray}
 &  &  &  & { 2.4} & $5.8 \times 10^{9}$ & $1.5 \times 10^{11}$ & 5.0 & 2000 & { 10.0} & { 10.0} \\ 
W~Aql & 2 epochs$^{\rm b}$ & -23.0 & 2.25 &  $5.9\pm0.3$ & $3.9 \times 10^{9}$ & $5\pm2 \times 10^{10}$ & 2.4 & 1000 & $1.5\pm1.0$ & $2.0\pm1.0$ \\ 
 &  &  &  &  $5.0\pm0.3$ & $5.5 \times 10^{9}$ & $4.5 \pm2\times 10^{10}$ & 4.5 & 1500 & $1.5\pm1.0$ & $2.0\pm1.0$ \\ 
 &  &  &  & $4.3\pm0.3$ & $7.4\times 10^{9}$ & $7\pm3 \times 10^{10}$ & 5.0 & 2000 & $1.5\pm1.0$ & $2.0\pm1.0$ \\ 
\rowcolor{Gray}
R~Aqr & 15-05-21 & -21.5 & 2.5 & { $4.63\pm0.03$} & $1.4 \times 10^{10}$ & { $4.3\pm0.5 \times 10^{10}$} & 11.3 & 1000 & { $4.0\pm0.2$} & { $-7.0\pm0.5$} \\ 
\rowcolor{Gray}
 &  &  &  & { $3.63\pm0.03$} & $2.3 \times 10^{10}$ & { $1.0\pm{0.1} \times 10^{11}$} & 12.6 & 1500 & { $4.0\pm0.2$} & { $-7.0\pm0.5$} \\ 
\rowcolor{Gray}
 &  &  &  & { $3.15\pm0.03$} & $\phantom{0}3.0 \times 10^{10}$ & { $3.5\pm0.3 \times10^{11}$} & 7.4 & 2000 & { $4.0\pm0.2$} & { $-7.0\pm0.5$} \\ 
$\pi$~Gru & 2 epochs & -11.0 & 1.45 & 2.5 & $2.2\times10^{9}$ & $5\times10^{10}$ & 2.6 & 1000 & 8.0 & 8.0 \\ 
&  &  &  & 2.0 & $3.4\times10^{9}$ & $1.2\times10^{11}$ & 2.7 & 1500 & 8.0 & 8.0 \\
&  &  &  & 1.8 & $4.2\times10^{9}$ & $2.0\times10^{11}$ & 2.9 & 2000 & 8.0 & 8.0 \\
\end{tabular}
\end{center}
\end{table}
\end{landscape}

\bibliographystyle{mnras}
\bibliography{../../bibliography}

\bsp	
\label{lastpage}

\end{document}